\documentstyle[prl,aps,epsf]{revtex} 
\input epsf 
 
\begin{document}

\title{Continuum Singularities of a Mean Field Theory of Collisions} 
 
\author{B.G. Giraud} 
\address{giraud@spht.saclay.cea.fr, 
Service de Physique Th\'eorique, DSM, CE Saclay,  
F-91191 Gif/Yvette, France} 
 
\author{and} 
 
\author{A. Weiguny} 
\address{weiguny@uni-muenster.de, 
Institut f\"ur Theoretische Physik, Universit\"at M\"unster, 
D-48149 M\"unster, Germany} 
 
\date{\today} 
\maketitle 
 
\begin{abstract} 
 
Consider a complex energy $z$ for a $N$-particle Hamiltonian $H$ and let  
$\chi$ be any wave packet accounting for any channel flux. The time  
independent mean field (TIMF) approximation of the inhomogeneous, linear  
equation $(z-H)|\Psi\rangle=|\chi\rangle$ consists in replacing  
$\Psi$ by a product or Slater determinant $\phi$ of single particle states  
$\varphi_i.$ This results, under the Schwinger variational principle, into  
self consistent TIMF equations  
$(\eta_i-h_i)|\varphi_i\rangle=|\chi_i\rangle$ in single particle space.  
The method is a generalization of the Hartree-Fock (HF) replacement of the  
$N$-body homogeneous linear equation $(E-H)|\Psi\rangle=0$ by single  
particle HF diagonalizations $(e_i-h_i)|\varphi_i\rangle=0.$  
We show how, despite strong nonlinearities in this mean field method,  
threshold singularities of the {\it inhomogeneous} TIMF equations are  
linked to solutions of the {\it homogeneous} HF equations. 
 
\end{abstract}

\section{Introduction} 
 
After the success of the mean field approach for bound state systems 
in various fields of physics, it was only natural to try the mean 
field concept for scattering states as well. The original attempt 
[1] was the time-dependent Hartree-Fock (TDHF) method where one solves  
the single-particle equations of motion as {\it initial} value problem in 
time. 
From the resulting solutions at various impact parameters one may then 
calculate 
the classical cross section. With no specification of the final state, 
the method is restricted to inclusive reactions. A serious, conceptual 
problem arises from spurious cross channel correlations [2,3]: when 
projecting the TDHF Slater determinant for large times on an orthogonal set 
of channel wave functions, the expansion coefficients and the respective 
S-matrix vary in time ad infinitum. To overcome the shortcomings of 
TDHF, the time-dependent mean field (TDMF) approach [2,3,4] expands the  
density in two sets of (bi-orthogonal) single-particle wave functions 
and solves the equations of motion as {\it{boundary}} value problem in time, 
fixing initial and final densities. It has been proven that for TDMF 
an S-matrix can be defined which becomes asymptotically constant [2]. The 
problem with TDMF lies in combining self-consistency with given boundary 
conditions in time [3,5]. No practicable algorithm for this highly 
``non-local'' problem exists upto date for use in actual numerical  
calculations. A third approach is the time-independent mean field  
(TIMF) method [6], based on a Schwinger-type variational principle [7] 
for matrix elements of the resolvent or T-operator between given initial 
and final states. The method uses two sets of variational single- 
particle functions, analogous to TDMF, and leads to inhomogeneous 
equations of Hartree-Fock type which can be solved iteratively for 
given total energy of the system. TIMF is free of the conceptual 
and practical problems of TDHF and TDMF, resp., and has been tested 
successfully on a number of simple systems. It can be extended to  
incorporate particle-hole correlations, as has also been done for TDHF,  
within a generalized random-phase-approximation [8]. The present paper 
goes beyond the above problems [9] by studying the continuum singularities 
of this TIMF approach for collisions. 
    
Consider a finite number $N$ of particles. Factorized wave packets  
(shifted Gaussians in momentum representation for example) make an  
overcomplete basis in their Hilbert space of wavefunctions. Hence the  
calculation of a retarded Green's function amplitude
${\cal D} \equiv \langle  
\chi | (z-H)^{-1} | \chi \rangle,$ where {\it i}) $\chi$ is a product,  
$|\chi\rangle=\prod_{i=1}^N|\chi_i\rangle,$ and {\it ii}) each single  
particle wave function  $\chi_i$ is real rather than complex, makes a  
fully generic problem. Such factorization simplifications are not  
physically restrictive and help in the analysis of a mean field theory  
of collisions, the subject of this paper. 
 
For the sake of simplicity, we deal yet with spinless, distinct  
particles only and short range interactions $v_{ij}$ for the Hamiltonian  
$H=\sum_{i=1}^N p_i^2/(2m_i)+\sum_{i>j=1}^N v_{ij}.$ The case of  
identical particles can be treated later and, in the following, any  
reference to a Hartree method may be understood as a reference to a  
Hartree-Fock (HF) method if necessary. Again for simplicity, we consider  
the calculation of diagonal collision amplitudes only,  
$\langle \vec k |V(E^+-H)^{-1}V |\vec k \rangle,$ Born term subtracted.  
Generalizations to distinct prior and post interactions, $V,V',$ are kept  
for future work. The state $ | \vec k \rangle $ is taken as a plane wave  
of relative motion in any two cluster channel ground state and the product  
$|\chi\rangle \equiv V|\vec k\rangle$ is a square integrable state in the  
$N$-particle space. Finally $z$ is any complex number $E + i \Gamma,$ and  
the usual limit $E^+$ at the end of any calculation reads  
$\Gamma \rightarrow +0.$ 
 
It is trivial to use the Schwinger variational principle [7] and  
show that ${\cal D}$ is the stationary value of the functional 
\begin{equation} 
{\cal F} \equiv \frac {\langle\Psi'|\chi\rangle\langle\chi|\Psi\rangle}  
{\langle\Psi'|(z-H)|\Psi\rangle},  
\end{equation} 
under variations of $\Psi,\Psi'.$ The corresponding Euler-Lagrange  
equations read, with retarded boundary conditions and arbitrary norms and  
phases of $\Psi$ and $\Psi',$ 
\begin{equation} 
(z-H)|\Psi\rangle=|\chi\rangle, \ \ \langle\Psi'|(z-H)=\langle\chi|. 
\end{equation} 
 
The variational equations which occur in the time independent mean field  
(TIMF) [6] theory of collisions read, 
\begin{equation} 
(\eta_i-h_i)|\varphi_i\rangle=|\chi_i\rangle, \ \ \  
\langle\varphi'_i|(\eta_i-h_i)=\langle\chi_i|, \ \ \ i=1,\ ...\ ,N. 
\label{timf} 
\end{equation} 
They are obtained from Eq. (1) when $\chi,$ and the approximation $\phi,$  
resp.  $\phi',$ chosen for $\Psi,$ resp. $\Psi',$ are products of  
single particle orbitals, $\chi_i, \varphi_i, \varphi'_i,$ respectively.  
Such TIMF equations are very simple [6]. Except for  
a single particle density operator $\rho$ defined non diagonally as 
$\rho(\vec r\,',\vec r)=\sum_i\varphi_i(\vec r\,')\varphi_i'^*(\vec r),$  
they are just Hartree(-Fock) equations completed by a right hand side,  
representing the image of the channel in single particle space. In the  
following, ${\cal F}$ is restricted to such factorized source functions  
$\chi$ and trial functions $\phi,\phi'$ and will be labeled $F.$ A saddle  
value under such a restriction of $\phi,\phi'$ is not necessarily unique  
any more. It will be denoted by $D$ instead of ${\cal D}$ and may request  
an additional, identifying label. Now our claim is:  
{\it bound and unbound solutions of the usual Hartree(-Fock) equations}, 
\begin{equation} 
(e_i-h_i)|\varphi_i\rangle=0, 
\end{equation} 
{\it induce singularities of the one-body variational conditions, Eqs. (3)}. 
 
This reminds, naturally, of the strict connection between the  
singularities of the linear, inhomogeneous problem  
$(z-H)|\Psi\rangle=|\chi\rangle$ in the $N$-body space and the solutions of  
the linear, homogeneous Schr\"odinger equation $(E-H)|\Psi\rangle=0$ in the  
same space. Because of the nonlinear nature of Eqs. (3-4) in single particle 
space, our claim is not obvious, and will be qualified in this paper. 
 
Actually, in a previous paper [10], the claim was already  
substantiated in part : those energies $E_H,$ for which a bound  
Hartree(-Fock) solution $\phi_H$ is found, generate poles of the approximate  
amplitude $D$ provided by saddle points of the restriction $F.$ 
Furthermore $(z-E_H) D \rightarrow |\langle\chi|\phi_H\rangle|^2/ 
\langle\phi_H|\phi_H\rangle$ when $z \rightarrow E_H.$ Despite the  
nonlinearity of the approximation, such a residue at such a pole is almost  
expected. The analogy with the poles of ${\cal D}$ at exact eigenvalues  
for bound states is striking. We are now interested in a more difficult  
question, namely, is there a similar analogy at higher energies, when  
singularities of scattering and rearrangement collisions (thresholds, cuts) 
occur?  
 
\bigskip 
In Section II we briefly recall a very simple, soluble model [11], 
used earlier among several other models to validate $D$ as an approximation  
of ${\cal D}.$ The model is reintroduced for pedagogical reasons first, to  
illustrate a derivation of Eqs. (3). Then, and mainly, it is used to provide  
a complete investigation of singularities, for it boils down to manipulations 
of polynomials. In Section III we introduce an enriched model, exactly  
soluble too. Section IV contains a generalization and discussion of the  
results obtained in Sections II and III. Finally Section V contains our  
conclusion.

\section{First model, bare propagation, symmetric mean field, two-body  
threshold}

In this soluble model, there are only two one-dimensional particles with  
just their kinetic energies, and different masses $m_i=1/(2a_i),$ hence  
$H=a_1p_1^2+a_2p_2^2.$ While the inversion of $z-H$ is numerically trivial  
and allows a good validation [11] of the TIMF approximation, the  
formal expression of $(z-H)^{-1}$ in terms of one-body propagators  
$(\eta_1-a_1 p_1^2)^{-1}$ and $(\eta_2-a_2p_2^2)^{-1}$ is less trivial,  
as it demands a convolution. The TIMF method consists in replacing the  
convolution by just one product, namely 
\begin{equation} 
(z-a_1p_1^2-a_2p_2^2)^{-1}|\chi_1\chi_2\rangle\  
\propto\ (\eta_1-a_1p_1^2)^{-1}|\chi_1\rangle\  
        (\eta_2-a_2p_2^2)^{-1}|\chi_2\rangle\,.  
\end{equation} 
This comes from variations $\delta / \delta \varphi_i$ of the functional  
$F.$ An additional simplification results from a further remark : in those  
representations where $\chi$ and $H$ are real, one finds from Eqs. (2) that  
$|\Psi'\rangle=|\Psi^*\rangle,$ hence the possibility of just one trial  
function $\Psi$ if one uses a Euclidian $\large(\,|\,\large)$ rather than  
a Hermitian $\langle\,|\,\rangle$ metric, 
\begin{equation} 
{\cal F} \equiv \frac {\large(\Psi|\chi\large)\,\large(\chi|\Psi\large)} 
{\large(\Psi|(z-H)|\Psi\large)}=\frac {\large(\chi|\Psi\large)^2} 
{\large(\Psi|(z-H)|\Psi\large)}\,.  
\end{equation} 
For the present two particle model, the factorization of $\chi$ into two 
single particle wave packets with real wave functions $\chi_1,\chi_2$  
allows us to use the following form of $F,$  
\begin{equation} 
F= \frac {\large(\chi_1\chi_2|\varphi_1\varphi_2\large)^2} 
{\large(\varphi_1\varphi_2|\,\left(z-a_1p_1^2-a_2p^2\right)\,| 
\varphi_1\varphi_2\large)} 
= \frac {(\chi_1|\varphi_1)^2(\chi_2|\varphi_2)^2} 
{z(\varphi_1|\varphi_1)(\varphi_2|\varphi_2) 
-(\varphi_1|a_1p_1^2|\varphi_1)(\varphi_2|\varphi_2) 
-(\varphi_1|\varphi_1)(\varphi_2|a_2p_2^2|\varphi_2)}\,. 
\end{equation} 
We assume that $\chi_1,\chi_2$ are real in the momentum representation.  
The functional being insensitive to the norms and global phases of  
$\varphi_1,\varphi_2,$ elementary manipulations of  
$\delta F/\delta\varphi_i$ yield, in the same momentum representation, 
\begin{equation} 
\varphi_i(p)=\frac{\chi_i(p)}{\eta_i-a_ip^2}\,,\ \ \ i=1,2, 
\end{equation} 
with 
\begin{equation} 
\eta_i=z- 
\frac {\int dp\, \varphi_j^2(p)\, a_j p^2} {\int dp\, \varphi_j^2(p)} = 
z-\eta_j - \frac {\int dp\, \chi_j^2(p)(a_jp^2-\eta_j)^{-1}} 
                 {\int dp\, \chi_j^2(p)(a_jp^2-\eta_j)^{-2}}\,, 
\ \ \ i=1,2, \ \ \ {\rm and} \ \ \ j=1,2, \ \ \ {\rm and} \ \ \ j \ne i. 
\end{equation} 
It is convenient at this stage to define the integrals, 
\begin{equation} 
I_i=-(\chi_i|\varphi_i)=\int dp\, \frac {\chi_i^2(p)} {a_ip^2-\eta_i}\,, 
\ \ \ i=1,2, 
\end{equation} 
and notice that Eqs. (9) then read, 
\begin{equation} 
I_j\, \frac {d\eta_j}{dI_j}=z-\sum_{i=1}^2\eta_i,\ \ \ j=1,2. 
\end{equation} 
If furthermore one defines auxiliary variables $\omega_i$ by the  
conditions  
\begin{equation} 
\eta_i=a_i\omega_i^2, \ \ \ \Im \omega_i > 0,\ \ \ i=1,2, 
\end{equation} 
then it is useful to define $J_j \equiv a_j I_j.$ And Eqs. (11) become, 
\begin{equation} 
2\, a_j \omega_j J_j\, \frac {d\omega_j}{dJ_j}=  
z-\sum_{i=1}^2 a_i\,\omega_i^2, \ \ \ j=1,2, 
\end{equation} 
where a contour in the upper half plane of the complex variable $p$  
defines the integrals, 
\begin{equation} 
J_j=\int dp\, \frac {\chi_j^2(p)} {p^2-\omega_j^2}\,,\ \ \ j=1,2. 
\end{equation} 
The special cases $\Im \eta_j \rightarrow 0$ while $\Re \eta_j \ge 0$ define 
cuts in the complex $\eta_j$-plane. These correspond to  
$\Im \omega_j \rightarrow 0$ in the $\omega_j$-plane. 
 
\smallskip 
When the two particles are identical, it may be interesting to  
symmetrize and antisymmetrize Eqs. (13) as,  
\begin{equation} 
\sum_{i=1}^2 a_i\, \omega_i\, J_i\, \frac {d\omega_i}{dJ_i} =  
z-\sum_{i=1}^2 a_i\,\omega_i^2, 
\end{equation} 
and 
\begin{equation} 
a_1\, \omega_1\, J_1\, \frac {d\omega_1}{dJ_1} - 
a_2\, \omega_2\, J_2\, \frac {d\omega_2}{dJ_2} = 0 \, , 
\end{equation} 
and identify cases where the mean field might break their symmetry. But  
we shall keep the particles, and/or their channel wave packets, 
distinct for a while. 
 
A soluble model, involving only the manipulation of polynomials, is  
obtained if one chooses the forms of the wave packets as follows, 
\begin{equation} 
\chi_j(p)=\left[\frac {\gamma_j}  
{\,\pi\, [(p-K_j)^2+\gamma_j^2]\,} \right]^{1/2}, \ \ \ j=1,2, 
\end{equation} 
yielding the simple result, 
\begin{equation} 
J_j= \frac{i\gamma_j}{\omega_j[(\omega_j-K_j)^2+\gamma_j^2]} 
   + \frac   {1}     {         (K_j+i\gamma_j)^2-\omega_j^2}= 
\frac{-\omega_j-i\gamma_j} 
{\omega_j(\omega_j-K_j+i\gamma_j)(\omega_j+K_j+i\gamma_j)} \, . 
\end{equation} 
Resulting polynomial equations turn out to have a lower degree  
if $K_j=0,$ for then $J_j$ becomes $J_j=-1/[\omega_j(\omega_j+i\gamma_j)].$  
As will be found in this and the next Sections, two kinds of singularities 
emerge, {\it i}) ``physical'' ones, which essentially depend on $z$ and  
are not very sensitive to ``technical'' parameters $K_j,a_j,\gamma_j,$ and  
{\it ii}) ``technical'' singularities, more sensitive to such parameters.  
The analytical continuation provided across $\eta$ cuts [12] 
by this $\omega$ representation is clear. 
 
Once Eqs. (13) have been solved, the saddle point values of the functional  
read, using Eqs. (7-11), 
\begin{equation} 
D=\frac{(a_1\omega_1^2+a_2\omega_2^2-z)J_1J_2}{a_1a_2} \, . 
\end{equation} 
The search for singularities of $D$ as a function of the physical energy 
$z$ thus consists in eliminating $\omega_1,\omega_2$ between Eqs. (13) 
and Eq. (19). The former read, after elementary manipulations which 
take advantage of Eq. (18) when $K_1=K_2=0,$  
\begin{equation} 
2 a_1 x^2 y + a_1 \gamma_2 x^2 - a_2 \gamma_2 y^2 + 2 y z + \gamma_2 z =0,  
\ \ \  
2 a_2 y^2 x + a_2 \gamma_1 y^2 - a_1 \gamma_1 x^2 + 2 x z + \gamma_1 z =0,  
\label{eqxy} 
\end{equation} 
where it was convenient to set $\omega_1=i x,\ \Re x>0,$ and  
$\omega_2=i y,\ \Re y>0.$ Equivalently, if we scale $x$ and $ y$ into  
$x=\gamma_1 x'$ and $y=\gamma_2 y',$ respectively, the same equations read, 
\begin{equation} 
(A_1 x'^2 + z)\, (1+2y') - A_2 y'^2 = 0, \ \ \  
(A_2 y'^2 + z)\, (1+2x') - A_1 x'^2 = 0\,, 
\end{equation} 
with $A_1 \equiv a_1 \gamma_1^2$ and $A_2 \equiv a_2 \gamma_2^2.$  
An elimination of $y$ between Eqs. (\ref{eqxy}) gives, 
\begin{mathletters}  
\begin{eqnarray} 
&2& a_1^3 \gamma_1 x^7 - 
a_1^2 (4 z - a_1 \gamma_1^2 + a_2 \gamma_2^2) x^6 -  
a_1 z (8 z - a_1 \gamma_1^2 + 4 a_2 \gamma_2^2) x^4 - \nonumber \\  
&2& a_1 \gamma_1 z (3 z + a_2 \gamma_2^2) x^3 - 
z^2 (4 z + a_1 \gamma_1^2 + 4 a_2 \gamma_2^2) x^2 - 
4 z^2 \gamma_1 (z + a_2 \gamma_2 ^2) x - 
\gamma_1^2 z^2 (z+ a_2 \gamma_2^2) = 0, \label{res12} \\ 
&y&\, = -\, \frac{ \gamma_2\, ( a_1\, x^3 + 2\,x\,z+ \gamma_1\,z ) } 
{ ( 2\,x + \gamma_1 ) \, ( a_1\, x^2 + z )  } \ . \label{substy} 
\end{eqnarray}  
\end{mathletters} 
The multiplicity of solutions is thus $7,$ which raises a problem for the  
identification of a solution, if possible unique, which accounts for a  
physical approximation. 
 
In turn, if one inserts Eq. (18) into Eq. (19), one obtains  
$D$ as a rational function of $\omega_1,\omega_2,$ or $x,y$ as well. 
Upon taking advantage of Eq. (22b), this rational fraction reduces into 
a rational fraction of $x$ only, hence a polynomial relation between 
$D$ and $x,$ with degree $1$ for $D,$ 
\begin{eqnarray} 
&a_1^2& a_2 \gamma_2^2 x^3 (\gamma_1 + x)^2 (a_1 x^3 + \gamma_1 z +  
2 x z) D = 4 a_1^3 x^7 (x + \gamma_1) + a_1^2 (a_1 \gamma_1^2 +  
a_2 \gamma_2^2 + 12 z) x^6 +  
12 a_1^2 \gamma_1 z x^5 + a_1 z (3 a_1 \gamma_1^2 +  
\nonumber \\ 
&4& a_2 \gamma_2^2 + 12 z) x^4 + 2 a_1 \gamma_1 z  
(a_2 \gamma_2^2 + 6 z) x^3 + z^2 (3 a_1 \gamma_1^2 + 4 a_2  
\gamma_2^2 + 4 z) x^2 + 4 \gamma_1 z^2 (a_2 \gamma_2^2 + z) x +  
\gamma_1^2 z^2 (a_2 \gamma_2 ^2 + z). 
\end{eqnarray} 
The same result is obtained if one uses Eq. (7) instead of Eq. (19). 
 
An elimination of $x$ between Eq. (23) and Eq. (22a) finally gives a direct, 
polynomial condition relating $D$ and $z,$ 
\begin{eqnarray} 
&z_1^2 z_2^2 (z_1 + 1) (z_2 + 1) (z_1 + z_2 + 1) \bar D^7 -  
4 z_1 z_2 (z_1 + 1) (z_2 + 1) (z_1 z_2 - 4 z_1   - 4z_2  - 4 )\bar D^6 -  
4 [3 z_1^3 z_2^2 + 3 z_1^2 z_2^3 +  
\nonumber \\ 
    &(20 z_1^3 z_2 + 58 z_1^2 z_2^2 + 20 z_1 z_2^3)   +  
     (16 z_1^3 + 88 z_1^2 z_2 + 88 z_1 z_2^2 + 16 z_2^3)   +  
     (32 z_1^2 + 84 z_1 z_2 + 32 z_2^2)   + 16 (z_1+z_2)  ] \bar D^5 +  
\nonumber \\ 
&16 [3 z_1^2 z_2^2 + 39 (z_1^2 z_2 + z_1 z_2^2)   +  
(32 z_1^2 + 91 z_1 z_2 + 32 z_2^2)   + 48 (z_1+z_2)   + 16  ] \bar D^4 +  
\nonumber \\ 
&16 [3 z_1^2 z_2 + 3 z_1 z_2^2 - (8 z_1^2 + 91 z_1 z_2 + 8 z_2^2)   -  
 88 (z_1+z_2)   - 64  ] \bar D^3 +  
\nonumber \\ 
&192 [- z_1 z_2 + 4 (z_1+z_2)   + 8  ] \bar D^2 -  
64 (z_1 + z_2 + 16  ) \bar D + 256 = 0, 
\end{eqnarray} 
where $z_i=a_i \gamma_i^2/z$ and $D$ is scaled as $D=\bar D/z.$ The  
degree $7$ for $x$ in Eq. (22a) is correctly reflected here by the same  
degree for $D$ (and $\bar D$). Conversely, given an amplitude $D,$ the  
degree of the polynomial condition, Eq. (24), with respect to $z$ is $4.$ 
Hence there are $7$ approximate amplitudes offered by TIMF for each energy, 
while the inverse problem, ``given the TIMF amplitude, find the energy'', 
has 4 solutions. 
 
This model, although soluble, thus creates a complicated Riemann surface.  
Criteria are necessary to select one physical sheet, or physical pieces of  
sheets. Obvious candidates are the conditions $\Re x \ge 0,\Re y \ge 0$  
when Eqs. (20) are solved. Concerning Eq. (24), the very definition of  
${\cal D}$ demands that ${\cal D}$ be real and negative if $z$ is real and  
negative. When $z>0$ with a slight and positive imaginary part, then  
$\Im {\cal D}$ must be negative. Those roots $D$ which show the same  
properties should thus help the identification of suitable sheets. 
 
The argument is made much simpler if the ``technical'' parameters 
$a_1\gamma_1^2$ and $a_2\gamma_2^2$ are taken equal to some common value  
$\theta.$ This amounts, in some sense, to consider identical particles,  
although $a_1$ may still differ from $a_2.$ Then Eq. (24) factorizes as 
\begin{equation} 
\left[4 - 4 (\theta+z) D + \theta (\theta+z) D^2 \right]^2 
\left[16 + 8 (3 \theta - 4 z) D + 4 ( 3 \theta^2 + 10 \theta z + 4 z^2 )  
D^2 + \theta^2 (2 \theta + z) D^3 \right]=0. 
\end{equation} 
If $\theta$ is used as a unit for $z$ and similarly $1/\theta$ is used as  
a unit for $D$ this reads as well, 
\begin{mathletters}  
\begin{eqnarray} 
\left[4 - 4 (1+z) D + (1+z) D^2  \right] &=&0, \\ 
\left[16 + 8 (3 - 4 z) D + 4 (3 + 10 z + 4 z^2) D^2 + (2 + z)  D^3  
\right]&=&0. 
\end{eqnarray}  
\end{mathletters} 
The presence of a squared polynomial as the first factor in Eq. (25) 
reflects a ``symmetry breaking'' by the mean field approximation. Indeed, 
when analyzing the corresponding solutions of Eqs. (\ref{eqxy}), one finds  
that each pair of roots $\{x,y\},$ with $x \ne y,$ is accompanied by a pair  
$\{y,x\},$ generating the same value of $D.$ Such a degeneracy thus makes, 
out of 4 of all the 7 solutions for $\{x,y\},$  two distinct values 
for $D$. All told, $D$ then takes 5 distinct values. The remaining 3  
solutions account for the degree $3$ present in the second factor of Eq. (25).  
It is easy to verify that such 3 solutions are ``symmetric'', namely $x=y.$  
Notice that the symmetry breaking generates a rational inverse function, 
\begin{equation} 
z_{bk}=\frac{(D_{bk} - 2)^2}{D_{bk} (4 - D_{bk} ) }\,, 
\end{equation} 
while the symmetry conservation generates an equation of degree $2$ for $z.$ 
Since $z_{bk}$ must be counted twice, one recovers the 4 solutions of the  
inverse problem.  
 
It turns out that the ``symmetry breaking'' sector violates the double  
condition, $\Re x >0,$ $\Re y >0.$ Hence the properties of this sector  
are listed in an Appendix only. Turning now to the symmetric amplitude  
$D_{sy},$ the choice of a physical branch is reasonably easy, see  
Figs. 1-2. In Fig. 1, the lower half plane contains a loop acceptable as  
a physical candidate. We verified that $\Re x >0$ for this loop. Despite  
a suitable $\Re x >0$ if $z>0,$ the other branch in Fig. 1 is clearly not  
acceptable, for it contains values $D_{sy}$ with positive imaginary parts.  
Nor can one accept the third branch, seen in Fig. 2, despite its correct  
sign for $\Im D_{sy}.$ For it violates both the limit  
${\cal D} \rightarrow 0$ when $|z| \rightarrow \infty$ and  
the obvious condition ``$\Re {\cal D} < 0$ if $z$ is real and negative''. 
Furthermore $\Re x$ is found unsatisfactory for this third branch. 
 
\begin{figure}[htb] \centering 
\mbox{  \epsfysize=105mm 
         \epsffile{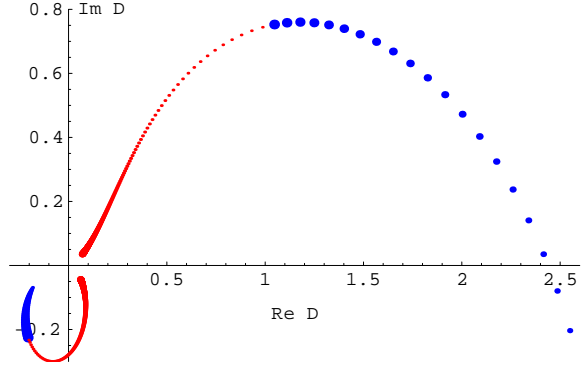} 
     } 
\caption{Complex $D$-plane. Two trajectories of symmetry conserving  
amplitudes as functions of $\Re z$ when $\Im z=1.$ Growing blue dots:  
$\Re z$ grows from $-\infty$ to $-0.$ Growing red dots: $\Re z$ increases  
from $+0.$ The third trajectory lies far in the lower half. Color is 
available online at www-spht.cea.fr/articles/t02/148/} 
\end{figure} 
 
Two values of $z$ generate branching for $D_{sy}.$ With a single root  
$D=-1/5,$ reasonable, a double root $D=16,$ unphysical, occurs for  
$z=-27/16,$ with expansion  
$D = 16 - 256/9\, (z+27/16) \pm 8192/243\, [-(z+27/16)^3]^{1/2}.$ Hence a  
familiar square root cut can be used to disentangle the two corresponding  
sheets, both unphysical. The value $z=-27/16$ does not represent a natural  
threshold for the present model. More physical, obviously, is the triple  
root singularity, $D=-2,$  which occurs at $z=0,$ the true threshold. It  
is illustrated by Fig. 3, where a tiny imaginary part $\Im z=.0001$ was  
added in order to separate branches. It will be noticed here that, 
although the physical branch gives real values of $D_{sy}$ when $z<0$ and 
complex values of the same when $z>0,$ there are always one real root and 
two complex conjugate roots on both sides in the vicinity of $z=0.$ This 
happens indeed because the corresponding discriminant, 
$\Delta_3=256z^2(27+16z)^3/(2+z)^4,$ actually changes sign, not for $z=0,$  
but rather for $z=-27/16.$ This helps to understand the nature of the  
unphysical singularity occurring at $z=-27/16.$ It gives an early  
``warning'' of the (cubic) physical threshold singularity, $z=0.$   
An elementary, but slightly tedious calculation provides  
the expansions of the 3 branches in the vicinity of $z=0,$ namely  
$D=- 2 - j\, 2^{2/3}\, 3\, z^{1/3} + {\cal O}(z^{2/3}) ,$ 
% - j^2\, 2^{1/3}\, 5\, z^{2/3} - 17z/3 ,$  
where $j$ is either $1,$ or any one of its complex cubic roots  
$\left(-1 \pm i\, \sqrt 3 \right)/2.$  
 
\begin{figure}[htb] \centering 
\mbox{  \epsfysize=105mm 
         \epsffile{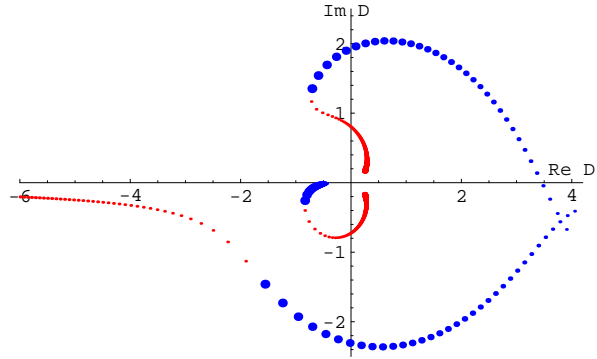} 
     } 
\caption{Complex $D$-plane. All trajectories $D_{sy}$ when $\Im z=.075.$  
Scales of trajectories made compatible by replacing radii from the origin  
by their square roots. Hence, for instance, announcing the double root $D=16$ 
when $z=-27/16,$ blue branches cross each other near $\sqrt D = 4.$ 
Growing blue dots: $\Re z$ grows from $-\infty$ to $-0.$ Growing red dots:  
$\Re z$ increases from $+0.$ } 
\end{figure} 
 
\begin{figure}[htb] \centering 
\mbox{  \epsfysize=105mm 
         \epsffile{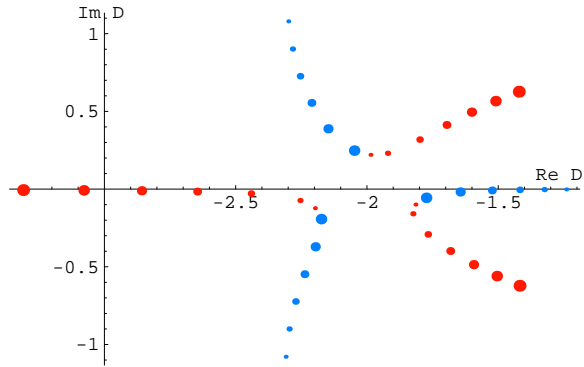} 
     } 
\caption{Complex $D$-plane. Triple merging $D_{sy} \rightarrow -2,$ with  
$\Im z=.0001,$  $-.09<\Re z<.09.$ Lower right branch physical.} 
\end{figure} 
 
Consider Eqs. (21) and set $A_1=A_2$ to factorize the 
resultant, Eq. (\ref{res12}). Then scale $x,$ $y$ and $z$ as proportional 
to $\gamma_1,$ $\gamma_2$ and $A_1,$ respectively. For the sake of simple  
numbers, this strictly amounts to set a common value 
$a_1=a_2=\gamma_1=\gamma_2=1,$ hence $A_1=A_2=1,$ for those reduced 
equations which govern the scaled variables and parameters.  
For the symmetry sector, 
$x=y,$ both equations, Eqs. (\ref{eqxy}), then boil down to  
$2 x^3 + 2 x z + z =0.$ It is trivial to  
find that, at threshold $z \rightarrow 0,$ all three roots have a leading  
term $x = (-z/2)^{1/3}+{\cal O}(z^{2/3}),$ while, as already found,  
$D_{sy}=-2+{\cal O}(z^{1/3}).$ Obviously, below threshold, 
one must select the real root $x,$ which gives a real amplitude. Conversely, 
above threshold, one must select that complex $x$ which gives a retarded 
amplitude. 
 
All told, for $D_{sy},$ cuts needed in the $z$-plane are a cut from  
$0$ to $+\infty$ for the cubic branching and, for instance, a ``technical'' 
cut from $-\infty$ to $-27/16$ to create an additional seam between the  
second and the third sheets. 
 
Now we consider additional cuts, namely those created by the condition  
$\Im \omega=0,$ or, identically, by the condition $\Re x=0.$ These occur  
because the solutions of realistic problems demand numerical, iterative  
calculations of $\eta_i$ and $\varphi_i$ before obtaining $D.$ This means  
inversions of operators $(a_i \omega_i^2 -h_i)$ in sequences of successive  
approximations of $\omega$'s (and self consistent $h$'s when potentials  
are involved). Obviously, every time $\Im \omega$ vanishes or becomes too  
small, numerical precautions are in order. Also, since the physical energy 
is on shell, $z=E+i 0^+,$ with a retardation boundary condition for  
many-body propagation, one would feel more comfortable with retardation  
also for the single particle energies $\eta \propto \omega^2.$ Advanced  
$\eta's$ are not to be ruled out {\it a priori}, because it is well known  
that mean field approximations can be excellent while breaking many-body  
symmetries. But, clearly, branches of $x$'s which cross such cuts  
$\Re x=0$ deserve some cautious scrutiny. 
 
For the present case where $A_1=A_2$ for ``symmetric'' bare propagations, 
and still with  simple numbers $a_i=\gamma_i=1,$ our results are shown 
in Figs. 4-5. (For the academic, ``symmetry breaking'' case, see the  
Appendix with Figs. \ref{cutxybk}-\ref{trjxybk}.) Fig. 4 is a contour 
plot of the product $\Re x_1\Re x_2\Re x_3$ of the real parts of the 3 roots 
as functions of $z$ in the $z$-plane. Darker areas indicate an increasing  
positive product (two out of the three $\Re x$'s are $<0$), while the 
lighter areas mean a more and more negative one (one negative $\Re x$ 
only). The product vanishes along the contour line separating the 
light grey area from the moderate grey one. It will be noticed that 
this line contains the point $z=0.$ Hence the cut relevant to $D$ and 
that relevant to $x$'s both contain the two-body threshold. Notice, 
however, that, except at such a treshold, a real $z$ induces complex  
$\eta$'s. Namely, propagation energy cuts {\it do not follow the real 
axis} in the $z$-plane.  
 
\begin{figure}[htb] \centering 
\mbox{  \epsfysize=100mm 
         \epsffile{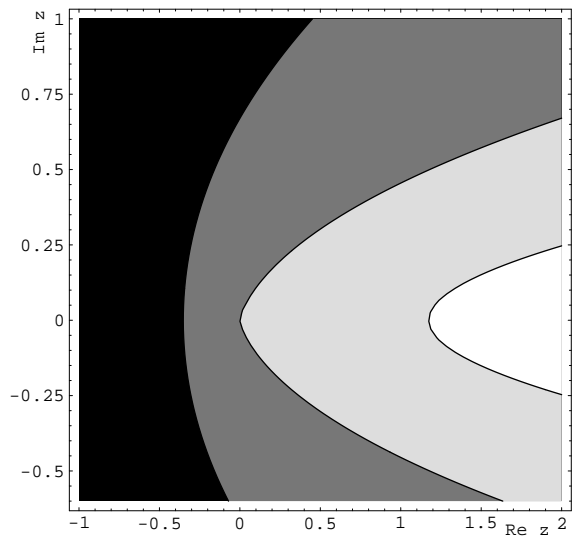} 
     } 
\caption{Complex $z$-plane. Cut caused by the condition $\Re x=0$  
for symmetry conserving roots. The cut is the contour line separating  
the lighter grey area from the darker grey one.} 
\end{figure} 
 
The next Figure, Fig. 5, shows the trajectories of the roots when we freeze  
$\Re z=0.1,$ above threshold, and let $\Im z$ run from $-1$ to $+1,$ hence  
allowing one $\Re x,$ then a second one, to change their signs. The sizes  
of dots are coded as follows: minimal for $\Im z=-1,$ growing until  
$\Im z=0,$ minimal again for small positive values of $\Im z,$ then growing 
again until $\Im z=1.$ The lower branch is the best candidate for physical  
roots, because it provides a growing retardation,  
$0< \Im \eta \equiv \Im (-x^2),$ when $\Im z$ is positive and grows. As  
predicted from Fig. 4, there is an interval for $\Im z$ where 2 roots $x$ 
have a positive $\Re x.$ 
 
\begin{figure}[htb] \centering 
\mbox{  \epsfysize=105mm 
         \epsffile{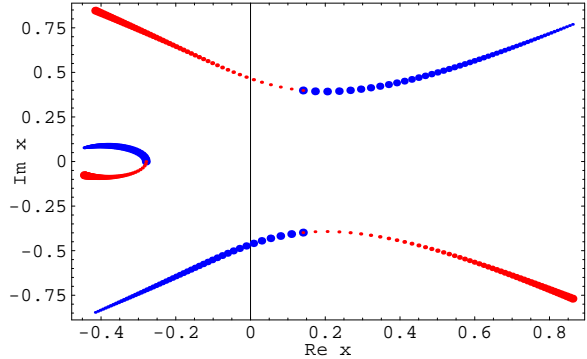} 
     } 
\caption{Complex $x$-plane. Trajectories of the symmetry conserving $x$'s 
when $\Re z=0.1,$ while $\Im z$ crosses the cut shown by Fig. 4. Blue dots  
growing when $\Im z$ grows from $-1$ to $0.$ Red dots growing when $\Im z$  
grows from $0$ to $1.$} 
\end{figure} 
 
To conclude this Section, the main result derived from this elementary 
model with bare propagation of two particles lies in the systematic,  
{\it physical}, two-body threshold found at $z=0$ in the energy plane  
($z$-plane) for all the mean field quantities, whether amplitudes $D$ or  
propagation energies $\eta.$ This threshold is, obviously, a common feature  
of both the exact problem and the corresponding Hartree problem. 
For amplitudes $D,$ a cut in the $z$-plane extends from the threshold  
$0$ to $+\infty,$ as seen in both the ``symmetric'' and ``breaking''  
submodels. For propagation energies $\eta,$ the cut starts from $z=0,$  
indeed, but  
deviates from the real semi-axis. For both $D$'s and $\eta$'s, the cost of  
the nonlinearity of the TIMF approach is reflected in additional, unphysical, 
``technical'' singularities. But such unphysical singularities are not  
beyond interpretation either, as shown by the analytical properties listed  
in this Section. Incidentally, as discussed earlier [13],  
unphysical singularities may be washed out by a linear admixture of the  
various solutions of the nonlinear mean field problem. The next Sections  
will show even better how physical cuts remain a significant feature of the  
TIMF approximation.

\section{Second soluble model, one-body threshold}

Here again we consider two one-dimensional particles, and particle 2 is  
still free with a pure kinetic energy $h_2=a_2p_2^2$ for its Hamiltonian.  
But now the complete Hamiltonian $H=h_1+h_2,$ while still separable,  
involves a bound state for particle 1, because we set 
$h_1=a_1p_1^2-\lambda |\chi_1\rangle\langle\chi_1|,$  
with an attractive enough potential. For technical reasons which will  
soon become clear, the form $\lambda |\chi_1\rangle\langle\chi_1|$ of this 
potential makes use of the same wave packet $\chi_1$ taken as a channel 
wave packet. The numerical inversion of $z-H$ is still easy and allows  
another good validation of the TIMF approximation. The formal expression  
of $(z-H)^{-1}$ in terms of one-body propagators $(\eta_1-h_1)^{-1}$  
and $(\eta_2-h_2)^{-1}$ demands again a convolution and the TIMF  
method consists in replacing the convolution by a product, 
\begin{equation} 
(z-h_1-h_2)^{-1}|\chi_1\chi_2\rangle\  
\propto\ (\eta_1-h_1)^{-1}|\chi_1\rangle 
      \, (\eta_2-h_2)^{-1}|\chi_2\rangle.  
\end{equation} 
This comes again from variations $\delta / \delta \varphi_i$ of the  
functional $F.$ And a further remark can be repeated: in those  
representations where $\chi$ and $H$ real, we obtain 
$|\Psi'\rangle=|\Psi^*\rangle,$ see Eqs. (2). Hence the possibility of just  
one trial function $\phi$ under a Euclidian rather than a Hermitian metric,  
see Eq. (6). The factorization of $\chi$ into two real wave packets  
$\chi_1,\chi_2$ essentially retains Eq. (7), which actually becomes,  
\begin{equation} 
F= \frac {\large(\chi_1\chi_2|\varphi_1\varphi_2\large)^2} 
{\large(\varphi_1\varphi_2|\,(z-h_1-h_2)\,| 
\varphi_1\varphi_2\large)} 
= \frac {(\chi_1|\varphi_1)^2(\chi_2|\varphi_2)^2} 
{z(\varphi_1|\varphi_1)(\varphi_2|\varphi_2) 
-(\varphi_1|h_1|\varphi_1)(\varphi_2|\varphi_2) 
-(\varphi_1|\varphi_1)(\varphi_2|h_2|\varphi_2)}. 
\end{equation} 
We use the same $\chi_1,\chi_2,$ real in the momentum representation.  
The functional being always insensitive to the norms and global phases of  
$\varphi_1,\varphi_2,$ the same manipulations of  
$\delta F/\delta\varphi_i$ yield, in the same momentum representation, 
\begin{eqnarray} 
|\varphi_1)=(\eta_1-h_1)^{-1}|\chi_1), \  
(\eta_1-a_1p_1^2)|\varphi_1)=|\chi_1)- 
\lambda|\chi_1)(\chi_1|\varphi_1), \  
\varphi_1(p)&=&\frac{\chi_1(p)}{\eta_1-a_1p^2}  
\left[1-\lambda (\chi_1|\varphi_1) \right], \\ 
\varphi_2(p)&=&\frac{\chi_2(p)}{\eta_2-a_2p^2}, 
\end{eqnarray} 
where it is better, temporarily at least, to retain the factor  
$\nu=\left[ 1-\lambda (\chi_1|\varphi_1) \right]$ for $\varphi_1.$  
The same quantity $\nu,$ as will be seen shortly, cannot be discarded from  
the self consistency conditions of the pair $\eta_1,\eta_2,$ 
\begin{eqnarray} 
\eta_1&=&z- 
\frac {\int dp\, \varphi_2^2(p)\, a_2 p^2} {\int dp\, \varphi_2^2(p)} = 
z-\eta_2 - \frac {\int dp\, \chi_2^2(p)(a_2p^2-\eta_2)^{-1}} 
                 {\int dp\, \chi_2^2(p)(a_2p^2-\eta_2)^{-2}}, \\ 
\eta_2&=&z-\frac{ (\varphi_1|h_1|\varphi_1) } 
{ (\varphi_1|\varphi_1) }= 
z-\eta_1-\frac { (\varphi_1|(h_1-\eta_1)|\varphi_1) } 
{(\varphi_1|\varphi_1)} = 
z-\eta_1 - \frac {(\chi_1|(h_1-\eta_1)^{-1}|\chi_1)} 
                {(\chi_1|(h_1-\eta_1)^{-2}|\chi_1)}. 
\end{eqnarray} 
Indeed, it is necessary to consider the matrix element, 
\begin{equation} 
{\cal I}_1=(\chi_1|(h_1-\eta_1)^{-1}|\chi_1), 
\end{equation} 
and notice that Eqs. (32-33) become, 
\begin{equation} 
I_2 \, \frac { d\eta_2 }{ dI_2 }=z-\eta_1-\eta_2= 
- \frac{(\chi_2|\varphi_2)}{(\varphi_2|\varphi_2)}\, ,\ \ \ \  
{\cal I}_1\, \frac{ d\eta_1}{ d{\cal I}_1 }=z-\eta_1-\eta_2= 
- \frac{(\chi_1|\varphi_1)}{(\varphi_1|\varphi_1)}\, . 
\end{equation} 
The integrals $I_1,I_2$ were already defined by Eq. (10). Returning to  
${\cal I}_1,$ and to the factor $\nu$ which accounts for the separable  
potential present in $h_1,$ an elementary manipulation of Eq. (30) gives, 
\begin{equation} 
(\chi_1|\varphi_1)=-{\cal I}_1, \ \ \ \  
{\cal I}_1=\frac{I_1}{1-\lambda I_1}. 
\end{equation} 
Again we define auxiliary variables $\omega_i$ by Eq. (12) and integrals  
$J_j \equiv a_j I_j$ by contours in the upper half plane of the complex  
variable $p.$  
Then Eqs. (35) become 
\begin{equation} 
2\, a_2 \omega_2 J_2\, \frac {d\omega_2}{dJ_2} =  
z-a_1 \omega_1^2-a_2 \omega_2^2, \ \ \ \  
2\, a_1 \omega_1 {\cal I}_1\, \frac {d\omega_1}{d{\cal I}_1} =  
z-a_1 \omega_1^2-a_2 \omega_2^2. 
\end{equation} 
It will be recalled here that a (unique) bound state occurs for $h_1$ 
for any positive value of $\lambda,$ at an energy $\eta_0<0,$ defined 
by the well known condition, 
\begin{equation} 
\frac{1}{\lambda}=\int dp\, \frac{ \chi_1^2(p) }{ a_1 p^2-\eta_0 } 
= \frac{J_1(\omega_0)}{a_1}, 
\ \ \eta_0=a_1 \omega_0^2,\ \ \Re \omega_0=0,\ \ \Im \omega_0>0. 
\label{bs1} 
\end{equation} 
Indeed, the right hand side is monotonically increasing when $\eta_0$ 
runs from $-\infty$ to $0$ and the same r.h.s. diverges at $\eta_0=0,$  
see Eq. (18), because of our choice of a Lorentzian form for $\chi_1^2.$ 
Accordingly, an explicit form of Eq. (\ref{bs1}) is, 
\begin{equation} 
\lambda+a_1\omega_0(\omega_0+i \gamma_1)=0, 
\ \ \Re \omega_0=0,\ \ \Im \omega_0>0, 
\end{equation} 
or, in terms of $\eta_0,$ 
\begin{equation} 
(\eta_0+\lambda)^2 + a_1 \gamma_1^2\, \eta_0 =0, 
\label{bs12} 
\end{equation} 
with obvious scaling properties. (Indeed, if the scale is set by $\lambda$  
for instance, it is convenient to define $A_1=a_1 \gamma_1^2,$ and the  
relevant scales are, obviously, $A_1/\lambda$ and $\eta_0/\lambda.$) 
Threshold singularities are expected for Eqs. (32-33) when $z$ reaches  
the one-body threshold $\eta_0,$ besides the already found two-body  
threshold $z=0.$  
 
The saddle point value $D$ deduced from Eq. (29) reads, upon taking  
advantage of Eqs. (30-37), 
\begin{equation} 
D= (a_1\omega_1^2+a_2\omega_2^2-z)\, {\cal I}_1\, I_2 \,. 
\end{equation} 
This formula, Eq. (41), is an obvious generalisation of Eq. (19). In the  
same way as we did in the previous Section, we shall again  
eliminate $\omega_1$ and $\omega_2,$ or rather the strictly equivalent 
variables $x=-i \omega_1$ and $y=-i \omega_2,$ between Eqs. (37) and Eq.  
(41). It is then useful to define a parameter $A_2=a_2 \gamma_2^2,$ quite  
similar to $A_1$ and it is also easy to predict that the solution $D(z)$   
scales in terms of $A_1/\lambda,$ $A_2/\lambda,$ $z/\lambda$ and  
$\lambda D.$ The Lorentzian choice for $\chi_1,\chi_2,$ induces the  
following forms for Eqs. (37), when we replace $a_1,a_2$ by 
$A_1/\gamma_1^2,A_2/\gamma_2^2,$ respectively, 
\begin{mathletters} \begin{eqnarray} 
A_1 \gamma_2^2 x^2 + 2 A_1 \gamma_2 x^2 y - A_2 \gamma_1^2 y^2 +  
\gamma_1^2 \gamma_2^2 z + 2 \gamma_1^2 \gamma_2 y z &=& 0,  
\\ 
2 \gamma_1 \gamma_2^2 \lambda x + A_2 \gamma_1^2 y^2 + 2A_2 \gamma_1 xy^2 -  
A_1 \gamma_2^2 x^2 + \gamma_1^2\gamma_2^2 z + 2\gamma_1 \gamma_2^2 x z &=& 0, 
\\ 
\frac{y}{\gamma_2} =-\,\frac{ ( \lambda \gamma_1^2 + A_1 x^2)\,x +  
(\gamma_1 + 2 x) \gamma_1^2 \,z } 
    { (\gamma_1 + 2 x) (A_1 x^2 + \gamma_1^2 z) }\,. 
\end{eqnarray} \end{mathletters}  
These scale obviously in terms of $x/\gamma_1$ and $y/\gamma_2.$ It is then  
convenient to set $\gamma_1=\gamma_2=1$ in Eqs. (42).  
 
Simultaneously, under the same replacement of $a_1,a_2$ by 
$A_1/\gamma_1^2,A_2/\gamma_2^2,$ respectively, we can take advantage 
of Eqs. (36) and (18) (with $K_1=K_2=0$) to let Eq. (41) become, 
\begin{equation} 
[\, A_1 A_2 (x^2 y^2 +\gamma_2  x^2 y + \gamma_1 x y^2 + 
\gamma_1 \gamma_2 x y) - A_2 {\gamma_1}^2 (y + {\gamma_2}) \lambda y\, ]\,  
D\, + A_1 \gamma_2^2 x^2 + A_2 \gamma_1^2 y^2 + \gamma_1^2 \gamma_2^2 z =  
0. 
\end{equation}  
Set $\gamma_1=\gamma_2=1.$ The elimination of $x$ and $y$ between Eqs. (42-43)  
yields a degree 7 polynomial condition for $D,$ 
\begin{eqnarray} 
&{\cal P}&(D, z, A_1, A_2,\lambda) \equiv A_2^2\, (A_1 + 4 \lambda)^2\,  
\left[\, ( z + \lambda )^2 + A_1 z\, \right]\,  
[\, ( z + \lambda + A_2 )^2 + A_1 ( z + A_2 )\, ]\, D^7 -  
\nonumber \\ 
&4& A_2 (A_1 + 4 \lambda)\, \left[\, (z+\lambda)^2 + A_1 z\, \right]\, 
[A_1 A_2^2 + 5 A_1 A_2 \lambda + 8 A_2^2 \lambda + 12 A_2 \lambda^2 +  
4 \lambda^3 + (-3 A_1 A_2 - 4 A_2^2  +  
\nonumber \\ 
&4& A_1 \lambda + 4 A_2 \lambda + 4 \lambda^2) z -  
4 ( A_1 + 2 A_2 + \lambda) z^2 - 4 z^3]\, D^6 \ + \ \ ... \ \ - \  
64 z (A_1 + A_2 + 20 \lambda + 16 z)\, D \, + 256 z\, =\, 0\, , 
\label{solvante2} 
\end{eqnarray} 
which is too cumbersome to be listed here entirely. A factor 
$\left[(z+\lambda)^2+A_1z\right]$ forces its coefficients for both 
$D^7$ and $D^6$ to vanish when $z=\eta_0,$ see Eq. (\ref{bs12}). Hence  
two roots $D$ diverge at the expected one-body threshold. We also 
notice that for $z=0$ the two lowest degree coefficients of ${\cal P}$ 
vanish, hence a double root $D=0$ occurs. But, for the sake of simplicity 
in this Section, we shall not elaborate much on the exact nature of this  
two-body threshold singularity for this second model. Similarities with 
the behavior of the first model around $z=0$ are likely. In the following 
we rather study in some detail the singularity at $z=\eta_0.$ 
 
The degree $7$ for $D$ is familiar from the model of the previous  
Section. But the degree for $z$ is now $5$ rather than $4.$ We  
verified that the limit $\lambda \rightarrow 0$ factorizes 
${\cal P}(D, z, A_1, A_2,\lambda)$ into a factor $z$ and a polynomial 
with degree $4$ for $z.$ 
 
It is convenient to set special values for a numerical investigation, for  
instance $a_1=\gamma_1=A_1=\gamma_2=1$ and $a_2=A_2=\lambda=2.$ The full  
polynomial then reads,   
\begin{eqnarray} 
& {\cal P} &\  =\   
 3\ (1 + z)\ (4 + z)\ \left[\ 27\ (3 + z)\ (6 + z)\ D\ +\ 12\   
 ( - 108 -  9 z + 14 z^2 + 2 z^3 )\ \right]\ D^6 - 
\nonumber \\ 
  4 &(&-2808 +  477 z + 5984 z^2 + 3480 z^3 + 690 z^4 + 44 z^5\,)\, D^5 +  
  8  ( - 584 + 1760 z + 3323 z^2 + 1387 z^3 + 200 z^4 +  8 z^5)     D^4 -  
\nonumber \\ 
  8 &(&- 140 + 1195 z + 1471 z^2 +  420 z^3 +  32 z^4\,)\,          D^3 +  
 32  ( -   4 +  107 z +   82 z^2 +   12 z^3) D^2 - 16 z (43 + 16 z) D + 64 z 
. 
\label{bon2} 
\end{eqnarray} 
Here, the bound state lies at $\eta_0=-1$ with $\omega_0=i.$ The second 
solution, $\omega_0=-2i,\eta_0=-4,$ of Eqs. (39-40) violates the condition  
$\Re x >0,$ hence pertains to an unphysical sheet. 
 
The 7 trajectories shown in Fig. 6 are those of the roots of  
${\cal P},$ Eq. (\ref{bon2}), when $\Re z$ runs from $-7.5$ to $+4.$  
This range suffices here to obtain a reasonable estimate of the root  
behavior when the energy runs from $-\infty$ to $+\infty.$ For the sake  
of graphical convenience, a renormalization $D/(1+|D|)$ forces large $D$'s  
back to the trigonometric circle. Also a small imaginary part $\Im z=.2$  
is set to enforce the rule $\Im {\cal D} <0.$ Black dots (or lines when  
nearing dots fuse) correspond to $\Re z < -4.1.$ Green and red ones  
correspond to $-3.9<\Re z<-1.1$ and $-0.9<\Re z,$ respectively. Finally,  
blue and yellow ones investigate neighborhoods, $-4.1<\Re z<-3.9$ and  
$-1.1<\Re z<-0.9,$ of expected singularities at $z=-4$ and $z=-1,$  
respectively. It turns out that Fig. 6 does not yield much information out  
of such ``blue'' and ``yellow' segments, although it is clear that only two  
``loops'' satisfy both rules $ \Im D <0,\,  \forall \Re z$ and   
$\lim_{|z| \rightarrow \infty} D = 0.$  Clearly, for a thorough investigation 
of all branchings and divergences, we should eliminate $D$ between ${\cal P}$  
and its derivative $\partial {\cal P}/\partial D,$ then study the neigborhoods  
of all the roots of the obtained resultant, 
\begin{eqnarray} 
{\cal R} = &z& (1 + z)^2 (2 + z) (3 + z) (4 + z)^2 (6 + z) (32 + 7 z)^2  
\times 
\nonumber \\ 
( 2426112 &+& 17293824 z + 54026784 z^2 + 121209152 z^3 + 233641545 z^4 + 
   328920768 z^5 +  
\nonumber \\  
  307812074 z^6 &+& 191171112 z^7 + 79534245 z^8 +  
     21923392 z^9 + 3826944 z^{10} + 380928 z^{11} + 16384 z^{12}\,)^3\, . 
\end{eqnarray} 
This straightforward but lengthy task gives too cumbersome results 
to be published here, naturally. Still, it might be useful to compare  
Fig. 6 with a superposition of Figs. 2 and \ref{loopsbk}, keeping in  
mind that symmetry breaking double roots of the previous model will  
now be disentangled. Indeed, under the already mentioned two criteria,  
namely i) $D \rightarrow 0$ if $|z| \rightarrow \infty,$ and ii)  
$\Im D<0,$ only the ``tiny'' loops selected from Figs. 2 and \ref{loopsbk}  
survive. Letting $\Im z \rightarrow 0,$ we obtained graphical evidence  
that such two loops grow in such a way that their ``blue'' and ``yellow''  
segments show the diverging roots predicted from the factor $(1+z)(4+z)$ in  
front of $D^7$ and $D^6.$ With the same renormalization $D/(1+|D|),$ Fig. 7  
confirms that two branches only are compatible with rules i) and ii), when  
we freeze $\Re z=-1$ and let $\Im z >0$ run. One of the ``good'' candidate  
roots diverges for $z=-1,$ see the green segment in the lower left part of  
Fig. 7. The other ``good'' candidate, $D_1 \simeq .31-.21i,$ see the small  
green segment at the beginning of the smallest trajectory in the lower  
right angle of Fig. 7, is a simple root as a function of $z$ in this area,  
and deserves little comment. The diverging root, however, because of its  
quadratic branching, deserves a study of its reciprocal, $d \equiv D^{-1}.$  
We set $z=-1+Z$ and expand ${\cal P},$ Eq. (\ref{bon2}), at lowest orders  
with respect to $d$ and $Z,$ 
\begin{equation} 
d^7\, {\cal P}(d^{-1},Z-1,A_1=1,A_2=2,\lambda=2) =  
270\, (2 d^2 + 9 Z) + {\cal O}( Z\,d). 
\end{equation} 
The neglected term is of order $Z^{3/2},$ because, obviously, the leading 
order of the double root is $d=\pm3 (-Z)^{1/2}/\sqrt 2,$ real below 
and imaginary above threshold, respectively. A similar, straightforward 
argument for the vicinity of the additional, but unphysical threshold 
at $z=-4$ yields the leading order $d=\pm 3 (-4-z)^{1/2}/(2\sqrt2).$ 
 
\begin{figure}[htb] \centering 
\mbox{  \epsfysize=105mm 
         \epsffile{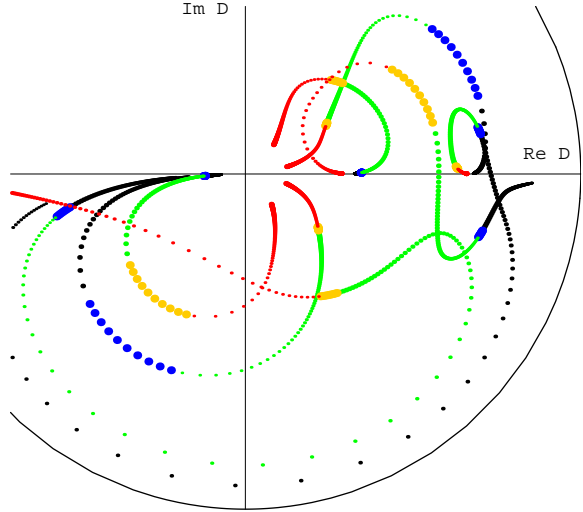} 
     } 
\caption{Complex $D$-plane. Trajectories of the 7 roots of Eq. (\ref{bon2}) 
when $-7.5 \le \Re z \le 4$ and $\Im z=.2.$ Black dots or lines correspond 
to $\Re z < -4.1.$ Blue, green, yellow and red ones correspond to 
$-4.1< \Re z < -3.9,$ $-3.9< \Re z < -1.1,$ $-1.1< \Re z < -0.9,$ and 
$-0.9< \Re z ,$ respectively. Only two trajectories always keep $\Im D <0$ 
and cancel $D$ when $|z| \rightarrow \infty.$} 
\end{figure} 
 
\begin{figure}[htb] \centering 
\mbox{  \epsfysize=105mm 
         \epsffile{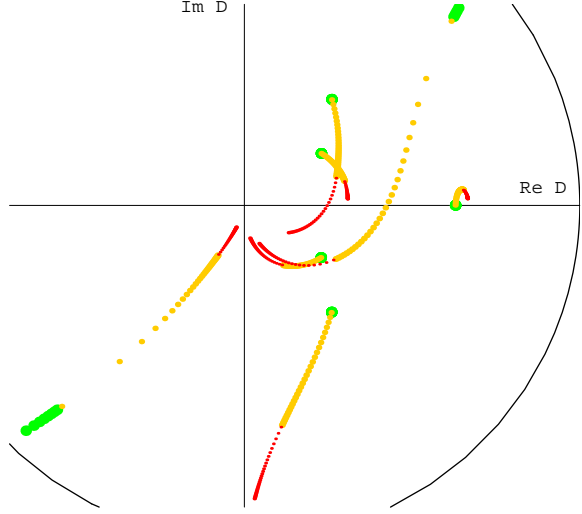} 
     } 
\caption{Complex $D$-plane. Trajectories of the 7 roots of Eq. (\ref{bon2}) 
if $\Re z=-1.$ Green, yellow and red mean $0 < \Im z < .01,$  
$.01< \Im z < 2$ and $2 < \Im z <8,$ respectively. Again, only two  
trajectories maintain $\Im D <0$ and cancel $D$ when  
$|z| \rightarrow \infty.$} 
\end{figure} 
 
Among all the singularities of this second model, we shall mainly discuss  
the physical threshold $z=-1.$ Eliminate $z$ between Eq. (42a) and Eq. (42b),  
or, equivalently, subtract the equations, Eqs. (37), from each other, hence 
\begin{equation} 
 a_2 \omega_2 J_2\, \frac {d\omega_2}{dJ_2}=  
 a_1 \omega_1 {\cal I}_1\, \frac {d\omega_1}{d{\cal I}_1}, 
\label{subtrac} 
\end{equation} 
a relation similar to Eq. (16). To prove the  
statement that the static HF energy $\eta_0$ indeed defines a  
threshold solution of the TIMF equations, Eqs. (37), it is enough to set  
$\eta_1 \rightarrow \eta_0,$ $\eta_2 \rightarrow 0$ and  
$z \rightarrow \eta_0.$ This automatically induces  
$z-a_1 \omega_1^2-a_2 \omega_2^2 \rightarrow 0,$ naturally. Set  
$\gamma_1=\gamma_2=1,$ for a trivial scaling. Then Eq. (\ref{subtrac}) reads, 
\begin{equation} 
\frac{      A_2\, y^2\, (1 + y)           } {1 + 2 y} = 
\frac{ x\, [A_1\, x\,   (1+ x) - \lambda] } {1 + 2 x} \,. 
\label{subtract} 
\end{equation} 
When $A_1=1$ and $A_2=\lambda=2,$ we know that the limits of interest are 
$x \rightarrow 1,$ $y \rightarrow 0$ and $z \rightarrow -1.$ These satisfy the  
condition, $\Re x >0,$ hence only $\Re y$ must be investigated. Define $X=x-1$ 
and $Z=z+1.$ Then Eq. (\ref{subtract}) boils down to $2 y^2=X,$ at  
leading orders in $y$ and $X.$ Accordingly, Eq. (42b), for instance, boils  
down to, $2y^2+Z=0.$ For $z<-1$ (below threshold) the solution,  
$y \rightarrow \sqrt{(-z-1)/2},$ is acceptable, with, simultaneously,  
$x \rightarrow -z.$ For $z>-1$ (above threshold), however, we find that 
a small, but positive $\Im z$ is necessary to allow the condition $\Re y>0.$  
This occurs because for $Z>0$ the leading order, $y^2 \rightarrow -Z/2,$  
actually generates $\Im y$ only. An expansion up to higher orders,  
is thus necessary for the knowledge of $\Re y.$ A straightforward, but  
slightly lengthy calculation yields, 
\begin{equation} 
y=i Z' - \frac{i}{3} Z^{\prime \, 3} - \frac{2}{3} Z^{\prime \, 4} +  
\frac{13\, i}{18} Z^{\prime \, 5} + {\cal O}\left(Z^{\prime \, 6}\right), 
\label{rac1} 
\end{equation} 
where $Z',$ a positive number, is defined as $Z'=\sqrt{Z/2}=\sqrt{(z+1)/2}.$ The ``formal conjugate'' of this expansion, 
\begin{equation} 
y= - i Z' + \frac{i}{3} Z^{\prime \, 3} - \frac{2}{3} Z^{\prime \, 4} -  
\frac{13\, i}{18} Z^{\prime \, 5} + {\cal O}\left(Z^{\prime \, 6}\right), 
\label{rac2} 
\end{equation} 
also holds, naturally. (Equivalently, it means the opposite choice of  
$Z',$ namely $Z'=-\sqrt{Z/2}.$) Both expansions induce a negative $\Re y$ as  
long as $Z$ is real and positive. The sign of this  $\Re y$ can be easily  
reversed, however, as soon as, above that threshold $z=-1,$ an imaginary  
part $\Im z$ is implemented. Another slightly cumbersome calculation defines, 
upon taking advantage of either Eq. (\ref{rac1}) or Eq. (\ref{rac2}), the  
condition for the border at which one of such roots acquires a positive 
real part. This is illustrated by Fig. 8. Near to that threshold $z=-1,$ the 
leading orders of the border condition give,  
$(\Im z)^2=\frac{2}{9}(\Re z+1)^5.$ Other numerical values for the  
parameters $\gamma_i,A_i$ etc. modify the numerical analysis, naturally,  
but leave intact the conclusion, namely that $|\Im z|$ must have at least a  
non vanishing value above the threshold if one needs one of these two roots  
to be compatible with the condition, $\Re y >0.$ 
 
\begin{figure}[htb] \centering 
\mbox{  \epsfysize=70mm 
         \epsffile{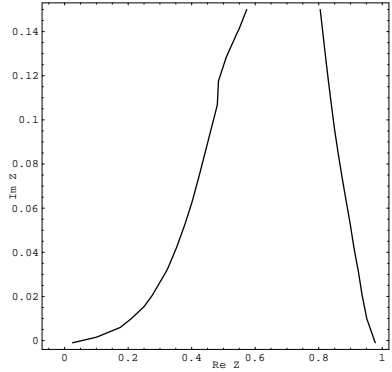} 
     } 
\caption{Complex $Z \equiv z+1$-plane. Below the plotted line, both roots  
described by Eqs. (\ref{rac1}-\ref{rac2}) show $\Re y<0.$ Above that line,  
one of them shows $\Re y>0.$ The line contains both thresholds $Z=0$ and  
$Z=1.$} 
\end{figure}

\begin{figure}[htb] \centering 
\mbox{  \epsfysize=105mm 
         \epsffile{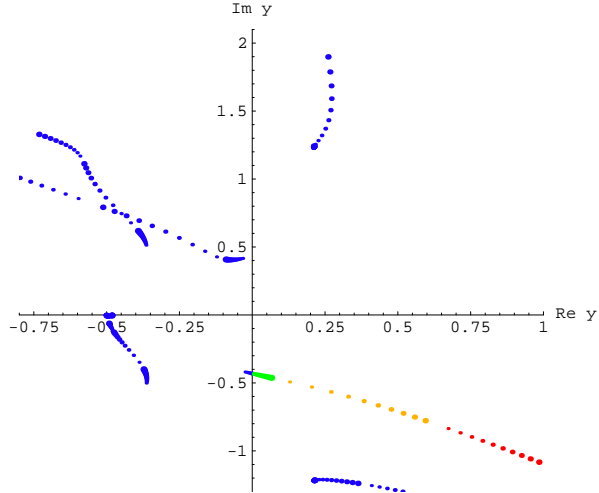} 
     } 
\caption{Complex $y$-plane. Trajectories of $y$ when $\Re z=-.6$ and $\Im z$ 
increases from $0.$ Blue lines are trajectories for which either 
$\Re x$ or $\Re y$ or both are negative. Only one branch, that long one 
in thelower right quadrant, survives the double  
condition, $\Re x >0,$ $\Re y >0.$ Tiny blue segment, $0 < \Im z <.03 .$  
Green segment,  $.04 < \Im z <.2 .$ Orange one,  $.3 < \Im z <1.8.$ Red one,   
$2 < \Im z <4 .$} 
\end{figure} 
We show on Figure 9 the trajectories of $y$ for $0<\Im z < \infty$  when  
$\Re z=-.6$ is frozen at an intermediate value between the thresholds  
$z=-1$ and $z=0.$  Only one branch is of interest, because all the other  
branches either stay in the $\Re y <0$ sector or the $x$ partner root shows  
$\Re x <0.$ The tiny blue segment at the beginning of this branch  
corresponds to $\Im z <.03,$ imaginary parts too small for letting  
$y$ acquire a positive real part, see Figure 8. 
 
\section{A theorem} 
 
We return to the case where $N$ is any finite particle number. The two-body  
interaction $V=\sum_{i>j}v_{ij}$ contained in the physical Hamiltonian $H$  
is assumed to be made of short ranged potentials $v_{ij}.$ Then the TIMF  
mean fields $U_i$  are also short ranged. For details of a further  
antisymmetrization with identical fermions, where the mean potential  
will be the same $U$  for all particles, we refer to [8];  
the short range of $U$ remains, whether one considers its direct or  
exchange part. At present we still retain the case of distinct particles.   
Eqs. (\ref{timf}) read again, 
\begin{equation} 
(\eta_i - t_i - U_i) |\varphi_i \rangle = | \chi_i \rangle, \ \ \  
\langle \varphi_i' |(\eta_i - t_i - U_i)= \langle \chi_i' | 
\,, 
\label{phi} 
\end{equation} 
with 
\begin{equation} 
\eta_i=z-\frac{\langle \phi' |H| \phi \rangle}{\langle \phi' | \phi \rangle}  
+ \frac{\langle\varphi'_i|(t_i+U_i)|\varphi_i\rangle} 
{\langle\varphi'_i|\varphi_i\rangle} 
\,. \label{eta} 
\end{equation} 
Notice that we now process a generalized argument, since we can also study  
non diagonal elements $\langle \chi'|(z-H)^{-1}|\chi \rangle$ where  
$\chi$ and $\chi'$ are products made of orbitals $\chi_i$ and 
$\chi_i',$ respectively. The Euclidian restriction is not 
implemented any more. The trial functions $\phi$ and $\phi'$ are the products  
made of orbitals $\varphi_i$ and $\varphi_i',$ respectively.  
All such quantities and wave functions depend on $z,$ but we stress here  
that, because of the short range of $U_i,$ the spectrum of $h_i=t_i+U_i$ has  
a fixed continuum, extending from $0$ to $+\infty$ on the real axis of the  
$\eta_i$ complex plane. In general $U_i$ is complex and the poles of  
$(\eta_i-h_i)^{-1}$ need not be real; as a matter of fact they move as  
functions of $z$ (and of the choices of $\chi$ and $\chi'$). But the  
continuum cut for the spectrum of $h_i$ remains always the same. It is  
therefore legitimate to ask the question ``what happens if one of the 
 $\eta_i$'s vanishes, hitting the threshold of the continuum of $h_i$?''.  
Incidentally, it will be noticed that there are many trajectories (sheets)  
of such $\eta_i$'s as functions of $z.$ The multiplicity comes not only from  
the existence of $N$ ``momenta'', $\omega_i \propto \pm \sqrt \eta_i,$ with  
their $\pm$ ambiguity [12], but it is also due to the nonlinearity of  
the mean field theory. For instance, in our second model, we found  
seven sheets, see the seven roots for each quantity $D(z),$ $x(z),$ $y(z)$ 
driven by $z.$ 
 
As a preliminary remark, we use Eqs. (\ref{eta}) to notice that the mismatch  
between any propagation energy $\eta_i$ and the corresponding self energy  
$ {\langle\varphi'_i|(t_i+U_i)|\varphi_i\rangle}/ 
{\langle\varphi'_i|\varphi_i\rangle} $ does not depend on $i.$ Furthermore  
we can take advantage of Eqs. (\ref{phi}) to relate the self and propagation 
energies as, 
\begin{equation} 
\frac{\langle\varphi'_i|(t_i+U_i)|\varphi_i\rangle} 
{\langle\varphi'_i|\varphi_i\rangle} = \eta_i + \frac{{\cal K}_i\,d\eta_i} 
{d{\cal K}_i}, \ \ \ \  
{\cal K}_i=\langle \chi_i' | (h_i-\eta_i)^{-1} | \chi_i \rangle = 
- \langle \varphi_i' | \chi_i \rangle = 
- \langle \chi_i' |\varphi_i \rangle \,. 
\label{misma} 
\end{equation} 
In other terms, the mismatch is measured by the ratio  
$\langle \chi'_i|\varphi_i \rangle / \langle \varphi'_i |\varphi_i \rangle = 
\langle \varphi'_i|\chi_i \rangle / \langle \varphi'_i |\varphi_i \rangle$ 
as a function of $\eta_i.$ When calculated at self consistent $\eta_i(z)$'s,  
such ratios do not depend on $i$ any more.

Assume that the special vanishing $\eta_s$ reads $\eta_s=i \varepsilon^2,$  
where $\varepsilon$ is a real, positive infinitesimal. This means that we  
select in the $z$ complex plane a trajectory which in turn induces an  
$\eta_s$ trajectory leading to retarded, outgoing boundary conditions for  
that special $\varphi_s$  and its partner $\varphi_s'.$  
For the sake of simplicity, set the inverse mass coefficient $a_s$ to unity, 
or, equivalently, renormalize $\eta_s$ and $U_s$ accordingly. In physical 
three dimensions, the partial wave components $\varphi_{s\ell}$ are described  
by differential equations of the form, 
\begin{equation} 
-\frac{d^2\varphi_{s\ell}}{dr^2}+\left[\frac{\ell(\ell+1)}{r^2}+U_{s\ell}(r)- 
i \varepsilon^2\right] \varphi_{s\ell}(r) = \chi_{s\ell}(r) - 
\sum_{\ell'}\int_0^{\infty}dr'\, U_{s\ell \ell'}(r,r')\, \varphi_{s\ell'}(r'), 
\ \ \ \ \forall \ell\,, 
\label{coupl} 
\end{equation} 
where $U_{s\ell \ell'}$ is a short notation accounting for, if necessary, 
partial wave coupling and/or non local parts of $U_s.$   
The source term $\chi_s$ is expanded in partial waves as well, naturally. 
It is then convenient to denote the right hand sides of Eqs. (\ref{coupl}) 
as source terms $\xi_{s \ell}(r).$ These are short ranged, obviously again.  
Similar equations hold for $\varphi'_s.$ 
 
For each $\ell$ let $\sigma_{s\ell}(r)$ be the regular solution, usually 
normalized as $\sigma_{s\ell}'(0)=1,$ of the homogeneous, left hand side of  
Eqs. (\ref{coupl}). The short range of $U_s,$ and similar short ranges  
assumed for $\chi$ and $\chi'$, make it that, when  
$r \rightarrow \infty,$ then $\varphi_{s\ell}(r)$ becomes  
$\simeq \exp[\,(i-1)\,\varepsilon\,r/\sqrt2\,]\, 
\int_0^{\infty} dr'\,\sigma_{s\ell}(r')\,\xi_{s \ell}(r'),$ 
with a similar asymptotic formula for $\varphi'_{s\ell}.$ Let $C,$   
a real and strictly positive number, be any convenient lower bound for  
the absolute values of these  
integrals $\int \sigma \xi$ and $\int \sigma \xi'$ in a neighborhood of 
$\varepsilon \rightarrow 0.$ This $C$ exists, since such integrals are  
usually finite and non vanishing when $\varepsilon=0.$ It is clear that,  
as $\varepsilon \rightarrow 0,$ there are no more any exponential decays  
or any asymptotic oscillations in the product $\varphi_s \varphi_s'.$ Then,  
at this limit for $\varepsilon,$ the integral  
$\langle \varphi_s' | \varphi_s \rangle$ diverges, while obviously an  
integral such as $\langle \chi_s' | \varphi_s \rangle$ remains  
finite. The ``mismatch'' cancels out. 
 
This indicates that, for any $i \ne s,$ the ratios 
${\cal K}_i\,d\eta_i/d{\cal K}_i$ vanish simultaneously at their respective  
energies $\eta_i.$ Besides  
threshold limits for each $\eta_i,$ there is an easy interpretation for such 
a situation, namely, each among such $N-1$ propagation energies converges  
towards a bound state energy of its $h_i.$ Indeed, let $de_i$ be an  
infinitesimal difference between $\eta_i$ and an isolated eigenvalue of $h_i.$ 
Then it is trivial, in an energy representation with biorthogonal eigenstates  
of $h_i,$ to see that ${\cal K}_i$ diverges at order $(de_i)^{-1},$ 
while $d{\cal K}_i/d\eta_i$ diverges at order $(de_i)^{-2}.$ 
 
The situation is thus representative of a Hartree(-Fock) solution for the  
$N-1$ particle system. This is confirmed by the observation that, since 
$\langle \varphi_s'|\varphi_s \rangle$ diverges, the potential $U_{i/s}$ 
induced by particle $s$ upon any particle $i \ne s$ vanishes. Indeed, 
the short range of $v$ in the formula, 
\begin{equation} 
U_{i/s}(r_i)=\frac{\int dr'\, v_{is}(r_i-r')\, \varphi_s'(r')\, \varphi_s(r')} 
{\langle \varphi_s'|\varphi_s \rangle}, 
\end{equation} 
makes the numerator converge $\forall r_i,$ while the denominator diverges. 
Any matrix element  
$\langle \varphi_i' \varphi_s'| v | \varphi_i \varphi_s \rangle$ will vanish  
too, for the same reason. Furthermore, the full matrix element  
$\langle \phi' |H| \phi \rangle/\langle \phi'|\phi \rangle$ can always be  
split as, 
\begin{equation} 
\frac{\langle \phi' |H| \phi \rangle}{\langle \phi' | \phi \rangle} =  
\frac{\langle \phi'_{-s} |H_{-s}| \phi_{-s} \rangle}  
{\langle \phi'_{-s} |\phi_{-s} \rangle} +  
\frac{\langle\varphi'_s|(t_s+U_s)|\varphi_s\rangle} 
{\langle\varphi'_s|\varphi_s\rangle}, 
\end{equation} 
where the subscript $-s$ refers to the subsystem where particle $s$ is  
removed. At the limit under study, both $\eta_s$ and  
${\cal K}_s\, d\eta_s/d{\cal K}_s$ vanish. Hence, according to  
Eq.(\ref{misma}), the self energy for particle $s$ vanishes and the full  
matrix element  
${\langle \phi' | H | \phi \rangle}/{\langle \phi' | \phi \rangle}$  
reduces to the subsystem value,  
${\langle \phi'_{-s} |H_{-s}| \phi_{-s} \rangle} / 
{\langle \phi'_{-s}|\phi_{-s} \rangle}.$ 
Furthermore, setting $i=s$ in Eq.(\ref{eta}), we find that $z \rightarrow  
{\langle \phi'_{-s} |H_{-s}| \phi_{-s} \rangle} / 
{\langle \phi'_{-s}|\phi_{-s} \rangle}.$ The threshold for the continuum  
of particle $s$ in the $z$-plane corresponds to the Hartree(-Fock) binding  
energy of the subsystem. 
 
Conversely, if $z$ converges towards a Hartree(-Fock) bound state energy of  
an $N-1$ particle system, it is easy to verify that at least one solution of  
the TIMF equations for the $N$ particle system consists in a threshold 
wave for the additional particle, as a spectator of the static solution for  
the subsystem. 
 
Notice that several special particles, not just one, can be forced into their  
continuum thresholds simultaneously. For instance, if particle $s$ and $s'$ 
are such that $\eta_s=\eta_{s'}=0,$ then all potentials $U_{i/s}$ and  
$U_{i/s'},$ including $U_{s/s'}$ and $U_{s'/s},$ vanish, and  
$z = {\langle \phi'_{-s-s'} |H_{-s-s'}| \phi_{-s-s'} \rangle} / 
{\langle \phi'_{-s-s'}|\phi_{-s-s'} \rangle},$ a subsystem energy for $N-2$ 
particles. 
 
It can be also noticed that such singularities do not depend upon the source  
terms $\chi$ and $\chi'.$ Indeed, the locations of such thresholds derive  
from homogeneous equations, where only $H$ appears.  
 
The present theorem can be phrased in a way which generalizes the theorem of  
[10]: not only the mean field binding energies of a system of  
$N$ particles define singularities of the TIMF propagator, but the mean field 
binding energies of its subsystems define thresholds of  
cuts where the additional particles become unbound.

\section{Discussion and Conclusion}

There are two parts in this work, namely on the one hand a couple of very 
special, analytical models, see Sections II and III, and on the other hand a  
theorem of a more general validity.  
 
\smallskip 
The systems described by our models are physically trivial, since they  
make non interacting particles. But their mathematical interest  
is different. As stated at the beginning of this work, it is important, for  
large particle numbers, to validate the replacement of convolutions by  
straight products, and our models allow a detail study of all singularities  
and nonlinearities introduced by the mean field approximation. We  
investigated three representations, namely what happens in, i) the $z$-plane  
(propagation energy), see for instance Figure 4, ii) the $D$-plane (TIMF  
amplitude), see for instance Figure 1, iii) pseudo momentum planes, such as,  
for instance the case of $x=-i \sqrt{\eta_1/a_1},$ see Figure 5. 
Since our models automatically implement an analytic continuation from  
physical to unphysical sheets, there is no cut to consider in the pseudo  
momentum complex planes. It is obvious, however, that for both pseudo momenta  
$x$ and $y$ the imaginary axis represents both rims of the cut which would  
be necessary in their respective $\eta$-plane. Accordingly, see for instance  
Figure 8, values of $z$ for which the real part of a pseudo momentum vanishes,
or identically for which a propagation energy $\eta(z)$ becomes real and  
positive, make cuts in the $z$ representation. The zoology of the TIMF  
solutions turns out to be surprisingly rich. The main two conclusions provided 
by the models can be listed as follows,  
 
\noindent 
i) except when the many-body propagation energy $z$ has too small 
an imaginary part, the TIMF equations always generate at least one branch of 
solutions where each single particle undergoes a retarded propagation 
and the TIMF amplitude $D$ shows all suitable properties needed for a  
reasonable approximation of a Green's function matrix element, 
 
\noindent 
ii) the threshold of a single particle continuum induces the threshold of a  
cut singularity in the $z$ representation; if one calls ``projectile'' 
that special particle becoming unbound, and ``target'' the system made by  
the other particle, the corresponding threshold value for the full  
propagation energy $z$ is the binding energy of the ``target''. 
 
\smallskip 
The theorem derived in Section IV, valid for any particle number $N \ge 2,$  
extends this numerical and analytical evidence. Hence the mean field theory 
of collisions mimicks the connection between singularities of the  
inhomogeneous problem $(z-H)|\Psi\rangle=|\chi\rangle$ and the solutions of  
the homogeneous Schr\"odinger equation $(E-H)|\Psi\rangle=0.$ At this 
stage of our work, the similarity is restricted, however: we considered 
only partitions where a ``target'' is surrounded by one or several unbound 
particles, and we have not proven thresholds defined 
by mean field energies of partitions $N_1+N_2=N,$ $N_1 \ge 2,$ $N_2 \ge 2,$  
into two clusters, each of them carrying its full internal energy.  
Nor have we considered even finer partitions $N_1+N_2+N_3=N,$ with 
$N_1 \ge 2,$ $N_2 \ge 2,$ $N_3 \ge 2,$ and so on. Last but not  
least, the present work lacks a clear description of the shapes of the cuts 
beyond their thresholds. The preliminary result obtained at the stage of  
Figure 8, with a ``border equation'' like $(\Im z)^2=\frac{2}{9}(\Re z+1)^5,$  
is an omen of subtle arguments yet to be phrased. 
 
\smallskip 
Despite such questions still open, the TIMF approximation now appears like a 
theory of collisions endowed with properties, such as poles and 
thresholds, with sound interpretations in terms of Hartree-(Fock) energies 
of subsystems. The special role played by single particle energy propagators 
$(\eta-h)^{-1}$ in the definition of such properties is a logical consequence 
of the factorization of trial wave functions, an essential ingredient of  
practical approximations. With the present and foregoing studies,  
TIMF appears as a reliable and practicable alternative to resonating 
group (RGM) or generator coordinate (GCM) studies for application 
in nuclear astrophysics where there is still a demand for microscopic 
rather than phenomenological calculations of processes relevant 
to element synthesis.

\bigskip \noindent 
Acknowledgement 
\smallskip \noindent 
A.W. thanks Service de Physique Th\'eorique, Saclay, for its hospitality  
during part of this work.

\section{Appendix: the symmetry breaking branch of the first model}

For the symmetry breaking sector of the first model, we again set   
$a_i=\gamma_i=1,$ scaling $x,y,z$ and ensuring the factorization of the  
resultant, Eq. (\ref{res12}). It is easy to analyze the singularities  
of the direct solution of Eq. (26a), 
\begin{equation} 
D_{bk}=2 \pm 2 \left(\frac{z_{bk}}{1+z_{bk}}\right)^{1/2}, 
\end{equation} 
in terms of one cut from $-1$ to $0$ in the complex $z$-plane, or,  
alternately, two cuts from $-\infty$ to $-1$ and from $0$ to $+\infty,$ and 
observe that the square root singularity at $z=0$ seems to represent a very 
traditional threshold singularity. Less physical, the r\^ole of $z=-1$ is  
to reflect the discriminant $\Delta_2=4z/(1+z)$ of Eq. (26a).  
 
\begin{figure}[htb] \centering 
\mbox{  \epsfysize=100mm 
         \epsffile{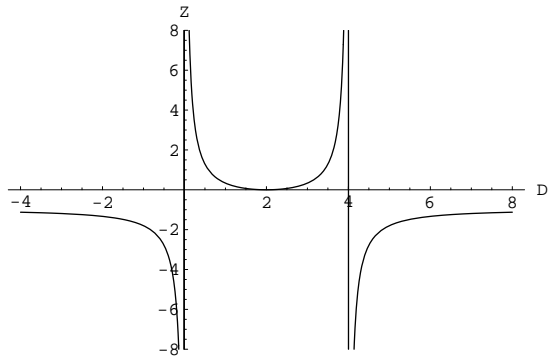} 
     } 
\caption{Scaled energy $z$ (unit $\theta$) as a function  
of the ``symmetry breaking'' amplitude $D$ (unit $1/\theta$).} 
\label{invbk} 
\end{figure}

In the forthcoming Figures, we keep $\theta=1,$ as a natural scale for  
energies and inverse amplitudes. Fig. \ref{invbk} shows the graph of $z_{bk}$  
when $D_{bk}$ is real and takes on all values from $-\infty$ to  
$+\infty.$ The symmetry axis at $D=2$ is obvious from Eq. (27). Since the  
physical amplitude is negative when $z$ is negative, the right lower  
branch of the graph is clearly unphysical, while the left lower branch  
is a reasonable candidate for approximations. (Notice, however, that no  
real estimate of the amplitude is offered for $-1 \le z \le 0.$) In turn,  
the right upper branch is also ruled out, as ${\cal D}$ must vanish when  
$z \rightarrow +\infty.$ This leaves the left upper branch as a tolerable  
candidate for physical approximates of the real (principal) 
part of ${\cal D}$ when $z$ is positive. 
 
\begin{figure}[htb] \centering 
\mbox{  \epsfysize=100mm 
         \epsffile{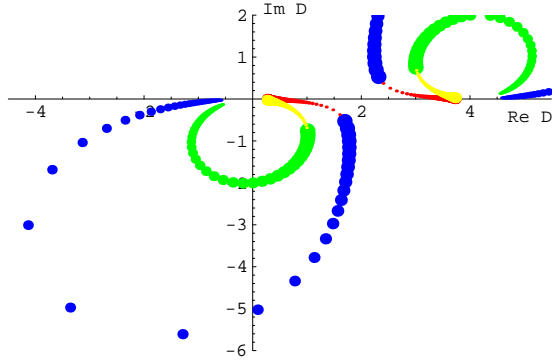} 
     } 
\caption{Complex $D$-plane. Trajectories of symmetry breaking $D$  
when $\Im z=.075$ (outer loops) and $\Im z=.4$ (inner ones). For  
$\Im z=.075,$ blue dots growing for $\Re z$ growing between $-\infty$ and  
$0^-$ and red ones growing for $\Re z$ growing from $0^+$ to $+\infty.$ 
For $\Im z=.4,$ blue and red replaced by green and yellow, respectively.} 
\label{loopsbk} 
\end{figure}

\begin{figure}[htb] \centering 
\mbox{  \epsfysize=100mm 
         \epsffile{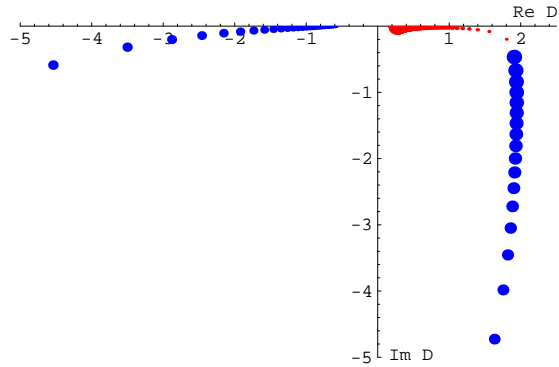} 
     } 
\caption{Complex $D$-plane. Lower loop trajectory of $D_{bk}$ when  
$\Im z=.02$ Blue dots growing for $\Re z$ growing between $-\infty$ and $0^-.$ 
Red ones growing for $\Re z$ growing from $0^+$ to $+\infty.$ } 
\label{anglbk} 
\end{figure}

Rather than considering inverse functions $z(D),$ we then show in  
Figs. \ref{loopsbk}-\ref{anglbk} the trajectories, in a complex plane  
``$D$'', of the solutions of Eq. (26a) when $\Re z$ takes on all values  
from $-\infty$ to $+\infty$ and $\Im z$ is frozen at some fixed value  
$\Gamma.$ The physical situation corresponds to $\Gamma=0^+,$ naturally,  
but Figs. \ref{loopsbk}-\ref{anglbk} use larger values of $\Gamma$ for  
graphical convenience. For Fig. \ref{loopsbk}, we use $\Gamma=.075$  
and $\Gamma=.4,$ which generate for $D_{bk}$ two ``outer loops'' and two  
``inner loops'', respectively. The r\^ole of $D=2$ as a symmetry center is 
obvious. The shrinking of the loops when $\Gamma$ increases comes from the  
fact that, as $|z| \rightarrow \infty,$ the dominant part of the symmetry  
breaking equation is $D(D-4)=0.$ Conversely, the evolution of such loops  
into ``angles'' when $\Gamma \rightarrow 0$ is transparent on  
Fig. \ref{anglbk}, obtained with $\Gamma=.02.$ 
\par \noindent 
Only those solutions which lie in the lower half plane can be retained  
as physical candidates, according to the condition ``if $\Im z >0,$  
then $\Im {\cal D}<0$.'' Hence the general physical behavior of $D_{bk}$  
is as follows:  
\par \noindent 
- when $z$ is real and increases from $- \infty $ to $-1,$ then $D$  
decreases from $0^-$ to $- \infty,$ 
\par \noindent 
- when $z$ is real and increases from $-1$ to $0,$ then $D$ varies from  
$2 - i \infty$ to $2,$ 
\par \noindent 
- when $\Im z=0^+$ and $\Re z$ increases from $0$ to $+\infty,$ then $D$  
decreases from $2 + i 0^-$ to $i 0^-.$ This infinitesimal imaginary part  
hints that $D_{bk}$ can at best approximate the principal part of  
${\cal D}.$ This was already deduced from Fig. \ref{invbk}. 
  
\begin{figure}[htb] \centering 
\mbox{  \epsfysize=100mm 
         \epsffile{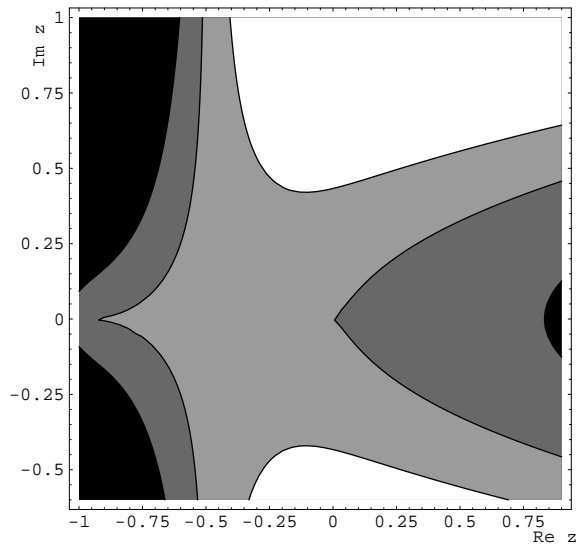} 
     } 
\caption{Complex $z$-plane. Cuts caused by the condition $\Re x=0$ for  
symmetry breaking roots. The cuts are the borders between light grey  
and darker grey areas.} 
\label{cutxybk} 
\end{figure}

The ``breaking'' factor of the factorizing resultant between Eqs.  
(\ref{eqxy}) reads, 
\begin{equation} 
x^4 - 2\, x^3\, z - x^2\, z - 2\, x\, z\, (1+z) - z\, (1+z) = 0, 
\label{brk} 
\end{equation} 
keeping in mind that its roots must be paired as $(x,y).$ Fig. \ref{cutxybk}  
displays a contour plot of the product of the corresponding 4 real parts  
of the roots as functions of $z.$ The corresponding cut in the $z$-plane  
is the border between the light grey and the darker grey areas. It is now  
made of two branches. The right hand branch, while not located on the  
real axis of the $z$-plane, again contains the two-body threshold $z=0.$  
 
\begin{figure}[htb] \centering 
\mbox{  \epsfysize=100mm 
         \epsffile{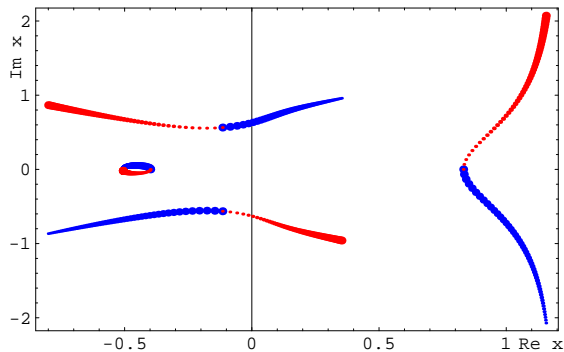} 
     } 
\caption{Complex $x$-plane. Trajectories of the symmetry breaking $x$'s  
when $\Re z=0.1,$  while $\Im z$ runs through one of the cuts shown by Fig.  
\ref{cutxybk}. Blue dots growing when $\Im z$ grows from $-1$ to $0.$  
Red dots growing when $\Im z$ grows from $0$ to $1.$} 
\label{trjxybk} 
\end{figure}

Then Fig. \ref{trjxybk} shows the trajectories of the 4 roots when we freeze  
$\Re z=0.1$ and let $\Im z$ run from $-1$ to $+1,$ allowing $z$ to cross  
twice the right hand side cut shown by Fig. \ref{cutxybk}. The sizes of the  
dots are coded like those of Fig. 5: minimal for $\Im z=-1,$ growing until  
$\Im z=0,$ minimal again for small positive values of $\Im z,$ growing again 
until $\Im z=1.$ The ``vertical'' branch on the right hand side of Fig.  
\ref{trjxybk} has the unsatisfactory property, that its ``$y$-partner''  
according to Eq. (\ref{substy}), is the loop like, tiny branch on the left  
hand side of Fig. \ref{trjxybk}. Hence $\Re x\, \Re y <0.$ In turn, the two  
``horizontal'' branches on Fig. \ref{trjxybk} are ``$x$-$y$'' partners and  
are partly located inside the right hand side of the complex $x,y$-plane.  
But actually the double condition, $\Re x>0,$ $\Re y>0,$  
is never satisfied. It must be concluded that the symmetry breaking sector  
is unphysical. 
 
A trivial manipulation of Eqs. (\ref{eqxy}) shows that the pairing of roots, 
for this sector and such special parameters, follows the rule, 
\begin{equation} 
x\,y=z-x-y,\ \ \ {\rm hence}\ \ \ y=\frac{z-x}{x+1}, 
\label{pair} 
\end{equation} 
which is its own inverse tranform, naturally. The rule is equivalent to  
Eq. (\ref{substy}), but simpler. Then if one defines $s \equiv x+y,$ it is  
easy to reduce Eq. (\ref{brk}) into, 
\begin{equation} 
(z-s)^2 = z(z+1),\ \ \ {\rm or}\ \ \ z-s= \pm [z(z+1)]^{1/2}, 
\end{equation} 
while shortening Eq. (\ref{pair}) into, 
\begin{equation} 
x\,y = z-s= \pm [z(z+1)]^{1/2}. 
\end{equation} 
This means that Eq. ({\ref{brk}}) factorizes into two distinct equations, 
\begin{mathletters} \begin{eqnarray} 
x^2 - x \left[z - (z+z^2)^{1/2}\right] + (z+z^2)^{1/2} =0\,, 
\label{eqp} \\ 
x^2 - x \left[z + (z+z^2)^{1/2}\right] - (z+z^2)^{1/2} =0\,. 
\label{eqm} 
\end{eqnarray} \end{mathletters} 
It is easy to verify that each of these is invariant under the transform, 
Eq. (\ref{pair}), hence each yields a pair $(x,y).$ A detailed analysis of  
all cases for such equations is trivial, but too lengthy to be published.  
Rather, it is enough and easy, actually, to set $\Im z=0,$ and plot, for  
instance for Eq. (\ref{eqp}), its two numerical roots as functions of $z.$  
It turns out that at least one of the roots has always a negative real part. 
The same phenomenon occurs for Eq. (\ref{eqm}). All told, the symmetry  
breaking sector does not respect the constraints requested simultaneously  
for $\Re x$ and $\Re y.$

\end{document}